\documentclass[10pt]{article}
\usepackage{amsmath}
\usepackage{amssymb}
\usepackage{graphicx}
\usepackage{cite}
\usepackage{color} 
\usepackage[labelfont=bf,labelsep=period,justification=raggedright]{caption}
\makeatletter
\renewcommand{\@biblabel}[1]{\quad#1.}
\makeatother
\date{}
\begin{document}
\begin{flushleft}
{\Large
\textbf{Determining superhydrophobic surfaces from an expanded Cassie Baxter equation describing simple wettability experiments}
}
\\
S. Kaps$^{1,\ast}$, 
R. Adelung$^{1}$, 
M. Scharnberg$^{1}$,
F. Faupel$^{2}$,
S. Milenkovic$^{3}$,
A.W. Hassel$^{3}$,
\\
\bf{1} Functional Nanomaterials, Technical Faculty, University of Kiel, Kiel, Germany
\\
\bf{2} Multicomponent Materials, Technical Faculty, University of Kiel, Kiel, Germany
\\
\bf{3} Institute for Chemical Technology of Inorganic Materials, Johannes Kepler University, Linz, Austria
\\
$\ast$ E-mail: ska@tf.uni-kiel.de
\end{flushleft}
\section*{Abstract}
The characterization of the wetting on superhydrophobic surfaces is rather complex. Usual contact angle experiments are difficult to perform and the lateral movement of droplets as well as the pinning at point defects on the surface can disturb the measurements. Even if precise contact angle measurements can be performed the information gain is limited if the surface is heterogeneously wetted. This results in the possibility of two surfaces with different roughness, different surface energy and thus different underlying wetting mechanisms exhibiting the same contact angle. We introduce the utilization of dynamic wetting experiments as an additional surface probe which allows a better characterization of superhydrophobic surfaces. A theoretical model is presented which describes the spreading of water jets on a superhydrophobic surface and allows the determination of the wetted fraction of a heterogeneously wetted superhydrophobic surface. The determined values for the wetted fraction identify a common problem when building artificial super hydrophobic surfaces and can fundamentally improve their understanding.
\section*{Introduction}
\begin{figure}[!ht]
\begin{center}
\includegraphics[width=6in]{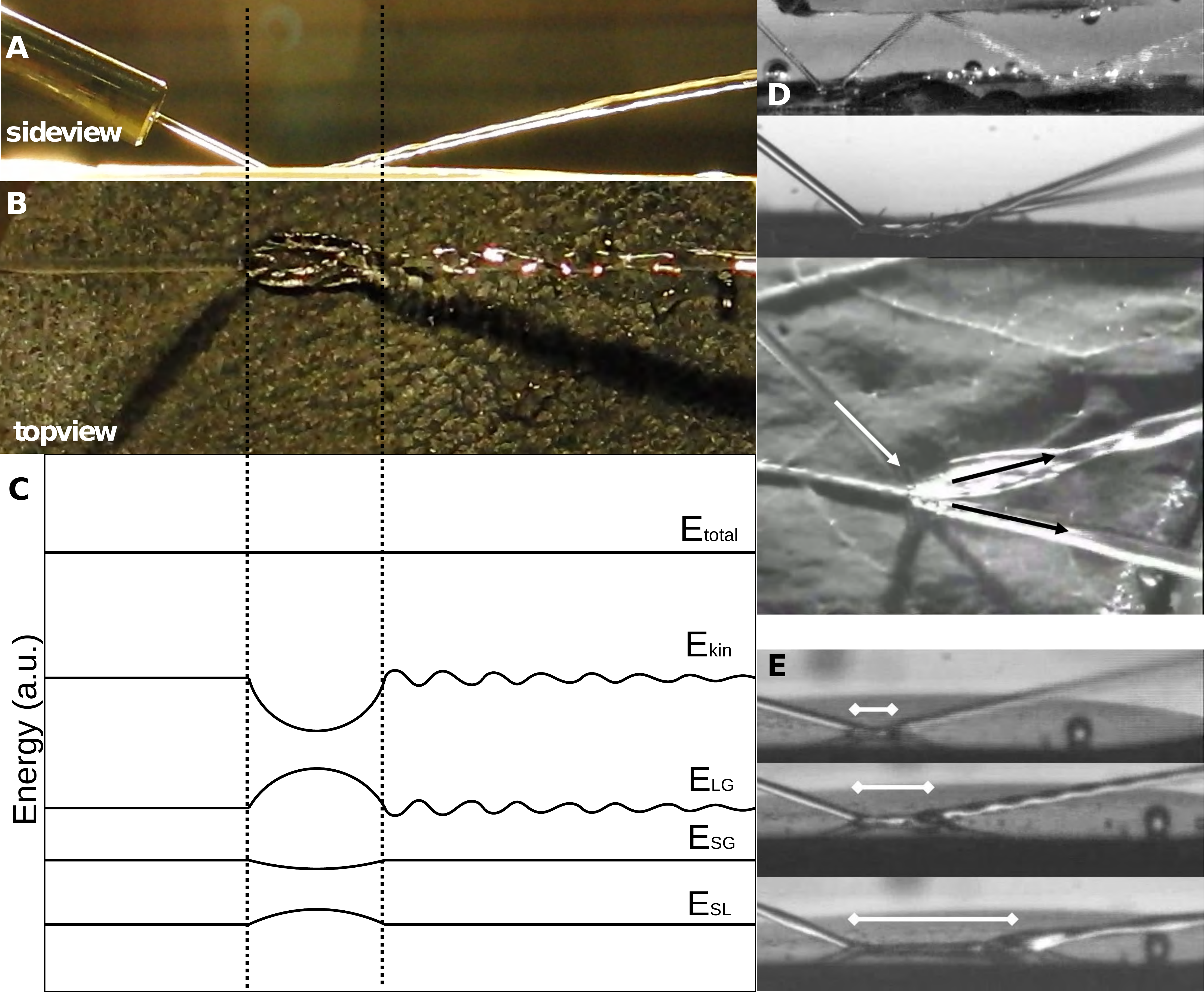}
\end{center}
\caption{
{\bf A water jet coming from the left side is reflected by a superhydrophobic surface.}  
\textbf{A} Sideview of a water jet which impacts on the surface, flows for a certain distance and lifts off as coherent jet.
\textbf{B} Topview showing the water jet spreading on the sample.
\textbf{C} Development of individual energetic terms through out the reflection. The distance correspond to the above illustrated sideview and topview images. The dashed lines indicate the point of impact and lift off. The kinetic energy $\text{E}_\text{kin}$ is transferred into new surface energy and the jet spreads. At the point of the maximum width the kinetic energy and the solid-gas interfacial energy ($\text{E}_\text{SG}$) reach a minimum whereas the solid liquid interfacial energy ($\text{E}_\text{SL}$) and liquid gas interfacial energy ($\text{E}_\text{LG}$) reach a maximum. Afterwards the jet contracts and lifts off the surface as a coherent jet with an elliptical cross section which causes some oscillation in $\text{E}_\text{kin}$ and $\text{E}_\text{LG}$.
\textbf{D} Complex reflection showing multiple reflections at two opposing surfaces and the splitting of one jet in two jets on a non-flat surface, e.g., leaf veins of nasturtium leaves.
\textbf{E} Influence of an increasing jet velocity on the reflection. The jet velocity is increasing from top to bottom, the arrows indicate the corresponding increase in flow length.}
\label{fig:waterjet_overview}
\end{figure}

The non-wetting of superhydrophobic surfaces is a useful phenomenon starting from the feathers of a duck ranging over functional outdoor clothes to self cleaning surfaces. Even though a general theoretical description was provided by Wenzel \cite{Wenzel1936} already in 1936 and expanded by Cassie and Baxter (CB) \cite{Cassie1945} in 1945 the experimental characterization of the wetting of such surfaces is a difficult task.  In the case of heterogeneous wetting as described by CB it is impossible to determine the roughness r and the wetted fraction f of the surface. Thus two samples with a different surface chemistry, a different roughness and a different wetted fraction can exhibit the same contact angle even though the underlying mechanisms of wetting differ. Therefore a contact angle measurement alone is not providing enough information to understand the wetting without further investigation. In this paper a theory is introduced which expands the theory of CB and makes use of a dynamic wetting phenomenon – water jet reflection – to improve the characterization quality of wetting experiments. Further it offers the possibility to distinguish between the wetting of two different samples which obey the same contact angle.
Hydrophobic materials with a certain roughness on a micrometer scale can exhibit contact angles larger than 160°. Wenzel \cite{Wenzel1936} described the wetting of surfaces with a certain roughness and Cassie and Baxter \cite{Cassie1945} introduced the term of heterogeneous wetting which implies that not the complete solid below a droplet is wetted but only a fraction. By measuring the contact angle infinite tuples of f and r can be found since the equation describing the heterogeneous wetting theory is under-determined. The combination of the high surface energy of water and the non wetting properties of superhydrophobic surfaces causes several interesting effects. Whereas a lateral movement of water droplets results in the Lotus effect \cite{Bhushan2009},\cite{Lee2006} a drop falling onto the surface can start bouncing \cite{Richard2002}. Recently Chiarot \cite{Chiarot2009} showed that a sequence of ink droplets can be deflected when shot on a superhydrophobic surface. For jets of liquids with a high surface tension there also exist some interesting phenomena such as the Kaye effect \cite{Kaye1963} and the deflection of mercury jets \cite{Baker1927}. Scharnberg \cite{Scharnberg2007} observed that a water jet that impacts a superhydrophobic surface the water spreading is followed by a contraction which causes a lift off and thus the water jet seems to be reflected by the surface. Figure \ref{fig:waterjet_overview}A shows the sideview of a typical water jet reflection on a superhydrophobic surface with a water jet coming from the left. It can be clearly seen that the points of impact and lift off are separated by several jet diameters due to a water flow on the surface. The topview, see figure \ref{fig:waterjet_overview}B shows the spreading of the water on the surface during the flow followed by a contraction towards the lift off. The energetic terms in figure \ref{fig:waterjet_overview}C sketch the individual energetic contributions throughout the reflection. The total energy $E_{total}$ must be preserved and is constant through the whole process. The kinetic energy $E_{kin}$ is transferred in interfacial energy indicated by the liquid-gas interface ($E_{LG}$), the solid-gas interface ($E_{SG}$) and the solid-liquid interface ($E_{SL}$) when spreading and back to kinetic energy when contracting. In certain cases an ellipsoid cross section of the jet after lift of occurs which can be described by the oscillation in $E_{kin}$ and $E_{LG}$ as depicted after lift off. The water jet reflection depends on parameters like surface geometry (see figure \ref{fig:waterjet_overview}D), jet velocity (see figure \ref{fig:waterjet_overview}E) and angle of incidence which affects the reflection in terms of angle of the reflected jet, flow length, spreading etc. In contrast to the reflection of light the underlying mechanism has a complex hydrodynamic nature and is not yet fully described. Celestini \cite{Celestini2010} developed a basic model which explains the fundamentals of the water jet reflection. The kinetic energy in the surface perpendicular velocity component of the water jet is transferred in surface energy, mainly $E_{LG}$, thus the water jet is spreading on the superhydrophobic surface. The spreading stops when the perpendicular velocity component becomes zero. Afterwards the surface energy is transferred back into the perpendicular velocity component with a directional shift of $180^\circ$ and the jet lifts off the surface.
\section*{Results and Discussion}
\begin{figure}[!ht]
\begin{center}
\includegraphics[width=3in]{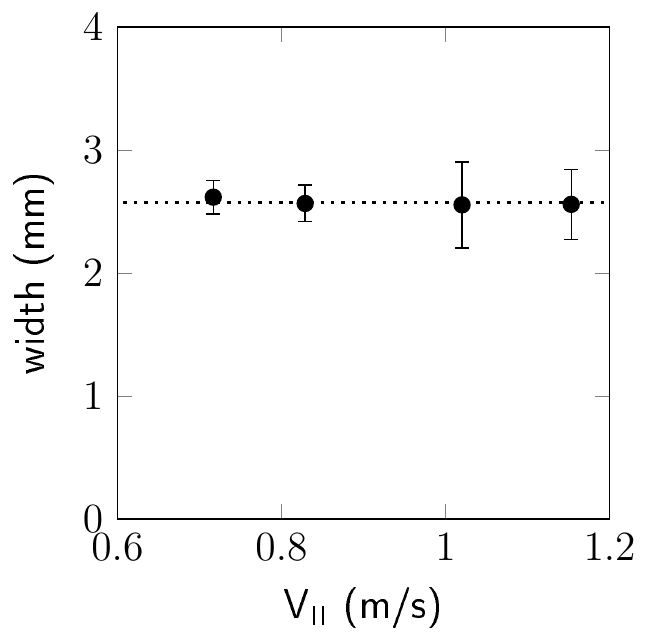}
\end{center}
\caption{
{\bf The maximum width of the water jet on the surface was measured for varying total jet velocities but with a constant perpendicular velocity component.}   It can be seen that the width is independent of the parallel velocity of the impinging water jet. 
}
\label{fig:width_vs_VII}
\end{figure}

\begin{figure}[!ht]
\begin{center}
\includegraphics[width=3in]{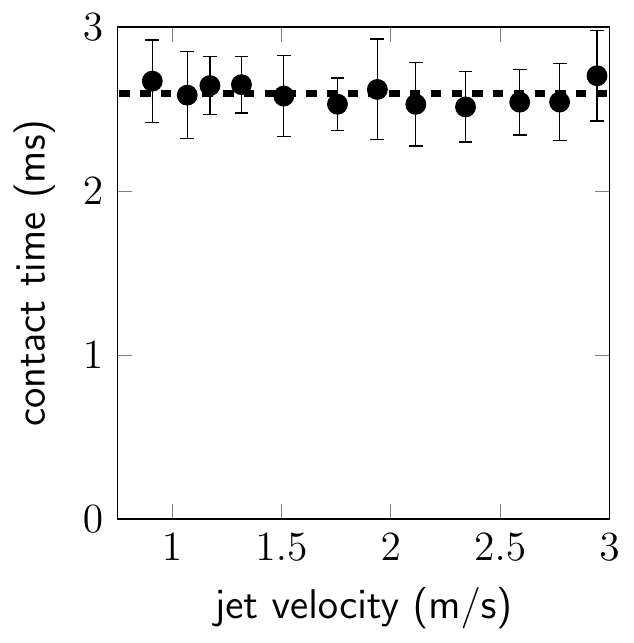}
\end{center}
\caption{
{\bf The contact time of the water jet with the surface can be calculated by measuring the flow length over the surface for a certain parallel jet velocity component.}   The contact time of the water jet was found to be independent of the jet velocity. 
}
\label{fig:contact_time}
\end{figure}

\begin{figure}[!ht]
\begin{center}
\includegraphics[width=2in]{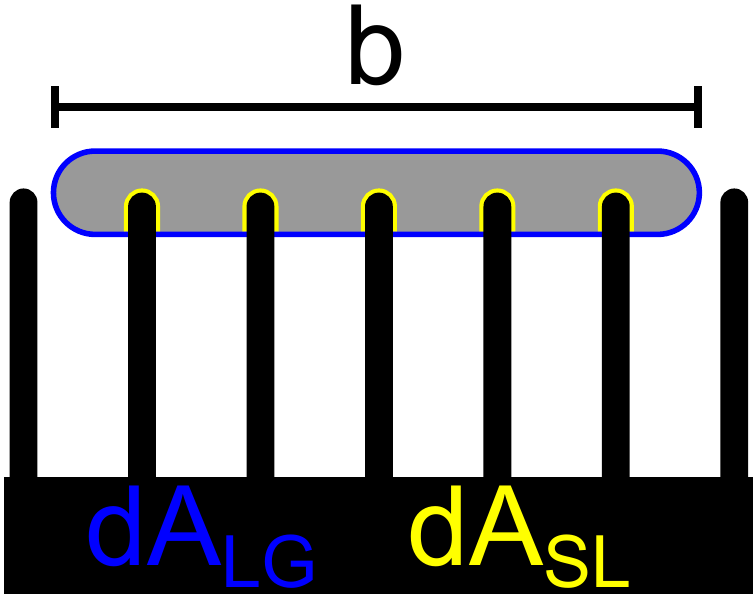}
\end{center}
\caption{
{\bf Schematic view of the water jet cross section at the point of maximum width b.}   The rough needle shaped surface is only partially wetting leading to a solid-liquid interface $\text{dA}_\text{SL}$, marked yellow. The air located between the needles leads to a liquid-gas interfacial area $\text{dA}_\text{LG}$, marked blue, with an undersurface size approximately equal the upper surface area.
}
\label{fig:Wetting_Schematics}
\end{figure}

\begin{figure}[!ht]
\begin{center}
\includegraphics[width=5in]{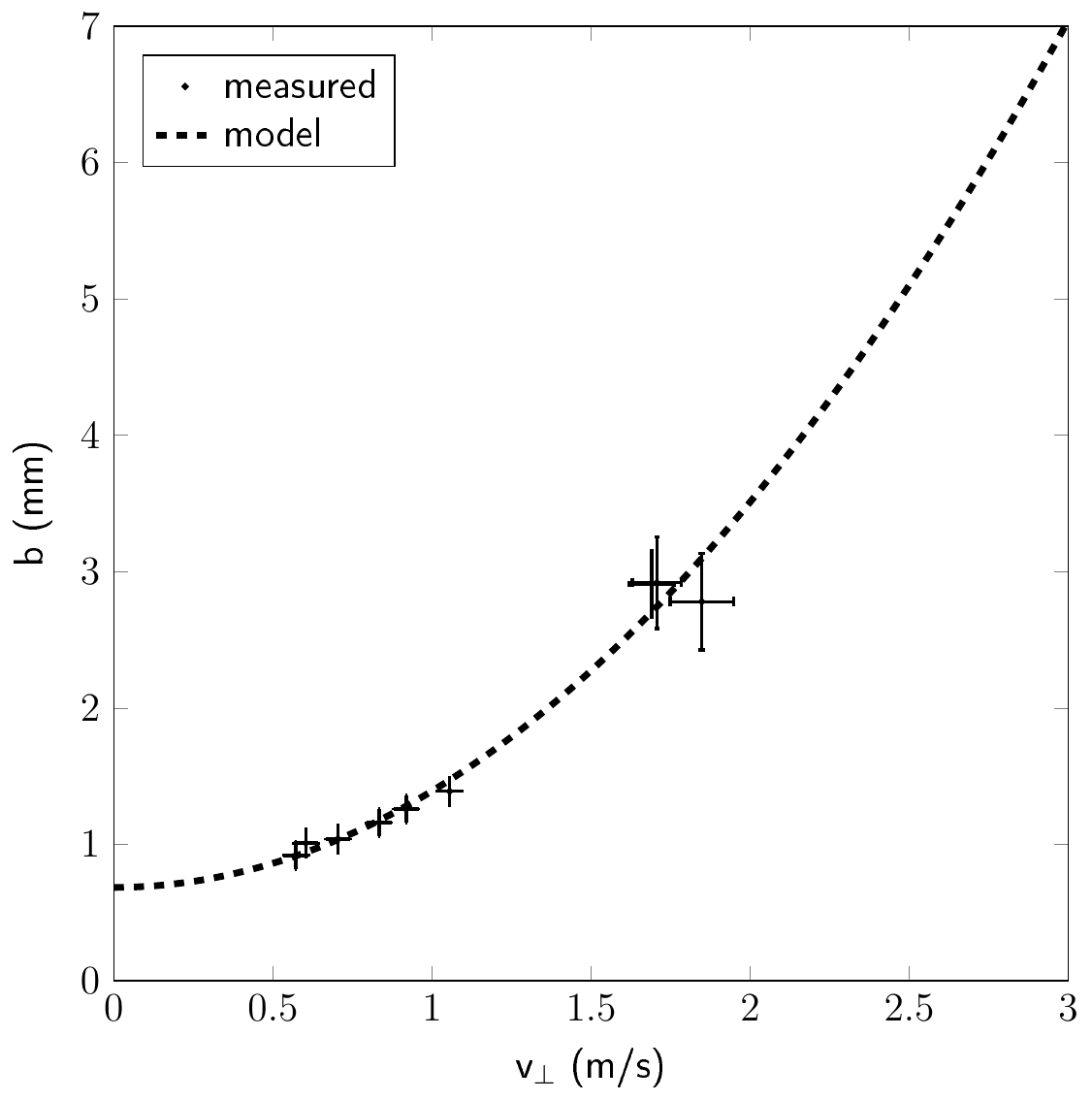}
\end{center}
\caption{
{\bf Maximum spreading width b for different velocities.}  The theoretical model matches the measured data.  The width was measured on Nasturtium leafs with a contact angle of $140.8^\circ$. Each data point is the average of 5 (low velocity) or 3 (high velocity) independent measured values. 
}
\label{fig:theory_reality}
\end{figure}

\begin{figure}[!ht]
\begin{center}
\includegraphics[width=2in]{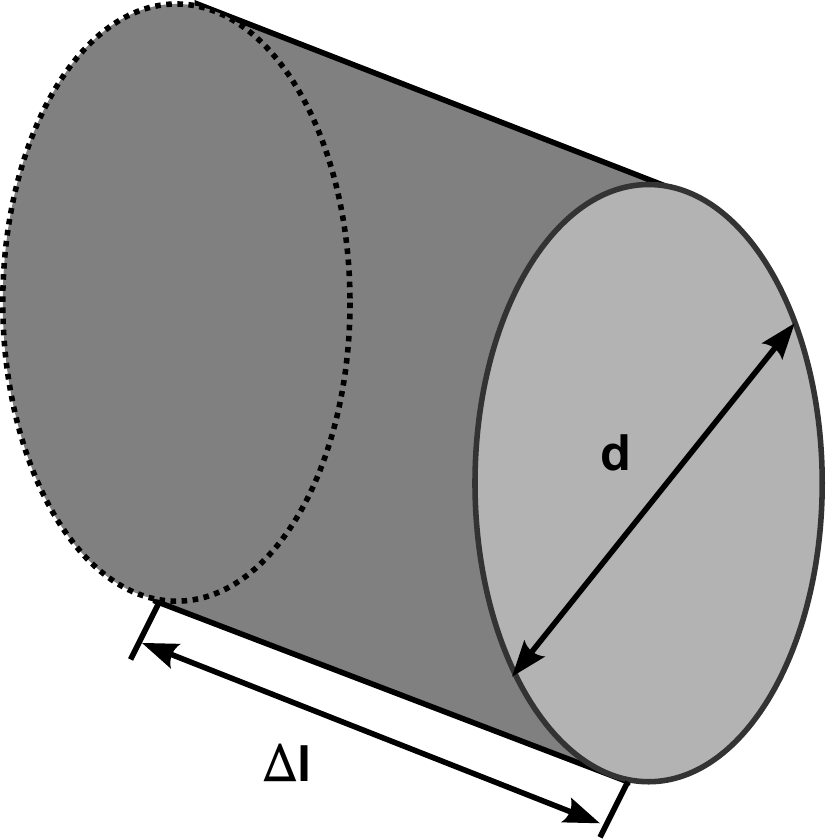}
\end{center}
\caption{
{\bf water jet shape}  Shape of the water jet before impact having a circular cross with diameter $d$ and length $\Delta l$
}
\label{fig:waterjet_shape}
\end{figure}

Here a theoretical model is introduced which bases on the conservation of energy during the reflection by comparing the energetic terms at two different well defined moments of the reflection: before hitting the surface and at the broadest point on the surface. The model is parameter free, directly derived from first principles and showing excellent agreement with experimental data. The details of the mathematical evolution of the individual equations can be found in the supplimentary part.
In order to expand common wetting theory all energies in solid-liquid and solid-gas interfaces were calculated with regard to the well established model by CB \cite{Cassie1945}, which introduces the heterogeneous wetting of a surface with a roughness $r$ and a wetted fraction $f$ below the droplet.
\begin{equation}
cos⁡(\Theta_{CB} )=rf\dfrac{\gamma_{SG}-\gamma_{SL}}{\gamma_{LG}} + f-1
\end{equation}
With $\gamma$ being the interfacial energy between solid-gas ($SG$), solid-liquid ($SL$) and liquid-gas ($LG$). The energy of the water jet before hitting the surface is the sum of the water surface energy and the kinetic energy of the jet.  Thus a jet as depicted in figure \ref{fig:waterjet_shape} with the velocity $v$ and the diameter $d$ has total energy per unit length $\Delta l$ of
\begin{equation}
\frac{E_0}{\Delta l}= \frac{1}{8} \rho\pi d^2 v^2+ \gamma_{LG} \pi d 
\label{eq:E_before}
\end{equation}
with $\rho$ being the density of the water. Other energetic terms as the potential energy and the viscoelastic energy can be neglected since the water jet reflection is only stable for small jet radii and non-turbulent water jets (in all experiments Raynold’s number $< 1000$). E.g. for the used radius of $300\,\mu m$ and a jet velocity of 1 m/s the potential energy is as low as 1\% of the kinetic energy.
For a negligible contribution of the viscosity the energy of maximum width can be calculated from the jet velocity and the interface energy since the energy must be conserved. Following the theory of bouncing droplets \cite{Adamson1990} the viscosity of the water only plays a minor role and can be neglected, if the contact time does not depend on the velocity of the droplet. The contact time of a water jet hitting a superhydrophobic surface was measured using a jet with a diameter of $600\,\mu m$ and an angle of incidence of $26.6^\circ$ with respect to the surface plane. Figure \ref{fig:width_vs_VII} shows the maximum width for surface parallel jet velocities ranging from $0.7\,m/s$ to $1.2\,m/s$ and a constant perpendicular velocity. It can be clearly seen, that the maximum width is not changing for a varying parallel velocity component. Therefore it can be stated that the kinetic energy in the parallel velocity component does not contribute to the spreading of the waterjet on the surface and is constant throughout the complete process. Following this finding the flow length $l_f$ (distance impact to lift off) is a direct measure for the contact time $t_c$, defined as
\begin{equation}
t_c = \frac{l_f}{v_{||}}.
\end{equation}
Figure \ref{fig:contact_time} shows the contact time for jet velocities ranging from $0.9\,m/s$ to $2.9\,m/s$ corresponding to perpendicular Weber numbers of $1.7$ and $5.4$ respectively. It can be seen that the contact time is constant over the whole measured region. This observation also agrees with the measurements performed by Celestini \cite{Celestini2010} for small perpendicular Weber numbers. Now, neglecting the viscosity of the water, a model of the water jet reflection can be developed which only bases on the principle of energy conservation. During the flow over the surface the kinetic energy is transferred into surface expansion. At the broadest point the perpendicular velocity component becomes zero and only the parallel velocity component is remaining. If the parallel component of the jet velocity is constant for the reflection, there is no influence on the maximum width and therefore the parallel velocity component does not play an important role for the reflection mechanism and can be neglected. By neglecting the parallel velocity component the kinetic energy at the point of the maximum width is zero, whereas the sum of the surface energies reaches its maximum. The total energy per unit length can be described as:
\begin{equation}
\frac{E_b}{\Delta l}= dA_{SL} \gamma_{SL}+dA_{LG} \gamma_{LG}+dA_{SG} \gamma_{SG}
\label{eq:E_interfaces}
\end{equation}
With $dA_x$ beeing the area per unit length. Following the theory of CB and using a wetting width b and a roughness of the liquid gas interface $R_{LG}$ equation \ref{eq:E_interfaces} can be expressed as:
\begin{equation}
\frac{E_b}{\Delta l}=b (rf(\gamma_{SL}-\gamma_{SG} )+R_{LG} \gamma_{LG} ) 
\end{equation}
By applying $v=v_\bot$ to equation \ref{eq:E_before} and the conservation of energy $E_0=E_b$ the wetted fraction is defined as
\begin{equation}
f = cos⁡(\Theta_{CB} )+1-R_{LG}+ \frac{\pi d}{b}\left(\frac{1}{8 \gamma_{LG}}\rho dv_\bot^2+1 \right).
\label{eq:wetted_fraction}
\end{equation}
Accordingly the maximum width is
\begin{equation}
b= \pi d \frac{1/(8 \gamma_{LG} ) d \rho v_\bot^2+1}{cos⁡(\Theta_{CB} )-f+1-R_{LG} }.
\end{equation}
The shape of the liquid-gas interface $R_{LG}$ is the only parameter which cannot be directly 
derived from common wetting theory. As an approximation the upper surface area of the water can be assumed to be flat, compare figure \ref{fig:Wetting_Schematics}. The curvature of the water-air interface is related to the pressure inside the water by the Laplace equation \cite{Adamson1990} leading to the same curvature on the undersurface \cite{Chiarot2009}. Since the undersurface and the upper surface have the same area the roughness of the liquid-gas interface can be expressed as $R_{LG}=2$.
To check the validity of the theory two different samples with a similar contact angle were selected. The experiments were carried out using artificial surfaces made from Re fiber arrays and natural ones made of leafs of the nasturtium. The surfaces exhibited contact angles of $140.8^\circ$ for the natural surfaces and $141.6^\circ$ for the Re fibers.
Water jet reflection experiments were carried out for both surfaces and the maximum spreading width was measured for the nasturtium leafs and for the artificial Re needles. By applying equation \ref{eq:wetted_fraction} the wetted fraction was calculated. Using a least square fit the calculated wetting fraction was found to be $f=0.950$ for the nasturtium leafs. It was observed that the theory agrees with the measurement over the whole measured region, see figure \ref{fig:theory_reality}, and the calculated wetted fraction is reasonable. Whereas the nasturtium leafs show a almost complete wetting the Re needles obey almost no wetting ($f=0.014$). This result perfectly fits the expectations since the Re needles have a very high aspect ratio and the wetting was expected to happen only at their tips leading to a low wetting fraction. Furthermore this result shows the superiority of nature, since it shows that a natural surface which was evolutionary optimized is perfectly suitable for water repellency without having mechanical instabilities. A wetted fraction of $f \approx 1$ presents an ideal surface in terms of super hydrophobicity since it obeys enough roughness to show a heterogeneous wetting with a high contact angle but is still flat enough to be mechanically stable. As shown in this work, artificial surfaces can obey similar contact angles but are not optimized for a high wetted fraction. This means that these surfaces contain a roughness which is unnecessary for a further improvement in terms of wetting but which mechanically destabilize the surface. In our experiments it was observed that the artificial surfaces are easily destroyed mechanically whereas the nasturtium leafs are extremely stable. In conclusion it can be stated that superhydrophobic surfaces or coatings should be optimized for a high wetted fraction which can be analyzed using the presented water jet reflection approach, which allows to determine f and r in the CB model only by wetting experiments for the first time.
\section*{Methods}
The artificial surfaces were formed by coating single crystalline Re fiber arrays with a thin hydrophobic polytetrafluoroethylen (PTFE) film \cite{Hassel2007}. A single crystal was grown from a eutectic NiAl-1.5 at.\% Re alloy by directional solidification in a Bridgman-type growth facility. Due to the strongly asymmetric composition of the eutectic alloy and the simultaneous solidification of both phases the minor phase (hexagonal Re) forms mechanical tough nanofibers in the major phase (B2-NiAl) \cite{Baker1927},\cite{Jackson1966},\cite{Philippe2007}. Temperature gradient and solidification speed can be used to control the resulting structure e.g. wire diameter and spacing \cite{Milenkovic2006} These nanofibers are all oriented in the same crystallographic orientation (isooriented) \cite{Hellawell1976} which was proven by XRD and EBSD \cite{Cimalla2008}. The major phase is removed by selective etching using a (HCl (32\%) : $\text{H}_2\text{O}_2$(30\%) : $\text{H}_2\text{O}$, 10 : 10 : 80) solution. The Re-nanofibers are not etched, yielding a sample with nanofibers standing upright on the alloy surface. By variation of the etching time, the fiber length can be precisely controlled. In order to obtain an artificial superhydrophobic surface, a PTFE thin film with a nominal thickness of 75 nm was deposited by sputtering \cite{SchA?rmann2005}.\par
The waterjet was generated using a nozzle with an inner diameter of $600\,\mu m$ which was connected to a closed water reservoir. The velocity of the waterjet was varied using different air pressures to generate a constant pressure inside the water container. The measurement of the maximum spreading width was obtained by extraction from photographs taken from above the surface, whereas side view photographs were taken to determine the angle of incidence. A very short exposure time of typically 5 ms was obtained using a Casio Exilim F1 digital camera. The contact time data was extracted from photographs taken from the side view which allows direct measurement of flow length $l_f$ and the angle of incidence. Each data point represents the average of typically 10 individual measurements. 
\section*{Acknowledgments}
We are grateful to acknowledge the German Science Foundation (DFG) for financial support under grant SPP 1165 (AD 183/3) HA and the Heisenberg Professorships (AD 183/5-2,).
\bibliography{Waterjet_Arxiv}
\section*{Figures}
\clearpage
\section*{Supplementary}
\paragraph{Mathematics I: Cassie Baxter}
Young's Equation with solid-liquid interfacial energy $\gamma_{SL}$, solid-gas interfacial energy $\gamma_{SG}$ and liquid-gas interfacial energy $\gamma_{LG}$ :
\begin{equation}
cos⁡(\Theta_{Y} )=\dfrac{\gamma_{SG}-\gamma_{SL} }{\gamma_{LG}}
\label{eq:Youngs_equation}
\end{equation}
Cassie Baxter Equation with wetted fraction $f$ and roughness $r$
\begin{equation}
cos⁡(\Theta_{CB} )=rfcos⁡(\Theta_{Y} )+ f-1
\label{eq:CB_basic}
\end{equation}
Cassie Baxter Equation after inserting Youngs equation
\begin{equation}
cos⁡(\Theta_{CB} )=rf\dfrac{\gamma_{SG}-\gamma_{SL}}{\gamma_{LG}} + f-1
\label{eq:CB}
\end{equation}
\paragraph{Mathematics II: Energy before impact}
Surface energy of a jet with surface area $A_0$
\begin{equation}
E_{S0} = A_0 \gamma_{LG} 
\end{equation}
Surface area for circular jet with diameter $d$ and length $\Delta l$
\begin{equation}
A_0 = \pi d \Delta l
\end{equation}
\begin{equation}
E_{S0} = \pi d \Delta l \gamma_{LG} 
\end{equation}
Kinetic energy of a jet with a mass $m$ moving with velocity $v$
\begin{equation}
E_{K0} = \frac{1}{2} m v^2 
\end{equation}
Mass of a jet with volume $V$ and density $\rho$
\begin{equation}
m = \rho V
\end{equation}
Volume of a jet with a circular cross section, diameter $d$ and length $\Delta l$
\begin{equation}
V = \pi\left(\frac{d}{2}\right)^2\Delta l
\end{equation}
\begin{equation}
E_{K0} = \frac{1}{8} \rho \pi d^2 \Delta l v^2
\end{equation}
Total energy before impact $E_0$
\begin{equation}
E_0 = E_{S0} + E_{K0}
\end{equation}
\begin{equation}
\frac{E_0}{\Delta l}= \frac{1}{8} \rho\pi d^2 v^2+ \gamma_{LG} \pi d 
\end{equation}
\paragraph{Mathematics III: Energy at broadest point}
Energy at the broadest point $E_b$ per unit length $\Delta l$ depends on the areas of the individual interfaces
\begin{equation}
\frac{E_b}{\Delta l}= dA_{SL} \gamma_{SL}+dA_{LG} \gamma_{LG}+dA_{SG} \gamma_{SG}
\label{eq:Eb}
\end{equation}
Following standard Cassie-Baxter theory:
\begin{equation}
dA_{SL} = b r f
\end{equation}
\begin{equation}
dA_{SG} = -b r f
\end{equation}
Liquid-Gas interface can be written as having a overall roughness $R_{LG}$
\begin{equation}
dA_{LG} = b R_{LG}
\end{equation}
Inserting in equation \ref{eq:Eb} yields
\begin{equation}
\frac{E_b}{\Delta l}= b r f \gamma_{SL} + b R_{LG} \gamma_{LG} - b r f \gamma_{SG}
\end{equation}
\begin{equation}
\frac{E_b}{\Delta l}=b (rf(\gamma_{SL}-\gamma_{SG} )+R_{LG} \gamma_{LG} ) 
\end{equation}
\paragraph{Mathematics IV: Water jet spreading}
By applying $E_0 = E_b$:
\begin{align}
b (rf(\gamma_{SL}-\gamma_{SG} )+R_{LG} \gamma_{LG} ) &= \frac{1}{8} \rho\pi d^2 v^2+ \gamma_{LG} \pi d\\
rf(\gamma_{SL}-\gamma_{SG} )+R_{LG} \gamma_{LG} &= \frac{1}{b}\left(\frac{1}{8} \rho\pi d^2 v^2+ \gamma_{LG} \pi d\right)\\
rf(\gamma_{SG}-\gamma_{SL} )-R_{LG} \gamma_{LG}&= -\frac{1}{b}\left(\frac{1}{8} \rho\pi d^2 v^2+ \gamma_{LG} \pi d\right)\\
rf\dfrac{\gamma_{SG}-\gamma_{SL}}{\gamma_{LG}}-R_{LG}&= -\frac{1}{b\gamma_{LG}}\left(\frac{1}{8} \rho\pi d^2 v^2+ \gamma_{LG} \pi d\right)\\
\underbrace{rf\dfrac{\gamma_{SG}-\gamma_{SL}}{\gamma_{LG}}+f-1}_{cos⁡(\Theta_{CB} )}-f+1-R_{LG} &= -\frac{1}{b\gamma_{LG}}\left(\frac{1}{8} \rho\pi d^2 v^2+ \gamma_{LG} \pi d\right)\\
cos⁡(\Theta_{CB} )-f+1-R_{LG} &= -\frac{1}{b\gamma_{LG}}\left(\frac{1}{8} \rho\pi d^2 v^2+ \gamma_{LG} \pi d\right)
\end{align}
The wetted fraction can be expressed as:
\begin{equation}
f = cos⁡(\Theta_{CB} )+1-R_{LG}+ \frac{\pi d}{b}\left(\frac{1}{8 \gamma_{LG}}\rho dv_\bot^2+1 \right)
\end{equation}
Accordingly the maximum width is:
\begin{equation}
b= \pi d \frac{1/(8 \gamma_{LG} ) d \rho v_\bot^2+1}{cos⁡(\Theta_{CB} )-f+1-R_{LG} } 
\end{equation}

\end{document}